\documentclass[twocolumn,floatfix,superscriptaddress,amsmath,showpacs,showkeys,aps,prb,longbibliography]{revtex4-2}
\usepackage[utf8]{inputenc}
\usepackage[final]{graphicx}
\usepackage{bm}
\usepackage[normalem]{ulem}
\usepackage{amsfonts}
\usepackage{amsmath}
\usepackage{amssymb}
\usepackage[usenames]{color}
\usepackage{times}
\usepackage{verbatim}
\usepackage{xcolor}
\usepackage{physics}
\usepackage{hyperref}
\usepackage{subfigure}

\begin{document}
\title{Laser induced $\mathcal{PT}$-symmetry breaking in the fluctuations of electronic fluids}
\author{Rui Aquino}
\affiliation{ICTP - South American Institute for Fundamental Research - Instituto de F\'isica Te\'orica da UNESP, 	Rua Dr. Bento Teobaldo Ferraz 271, 01140-070 S\~ao Paulo, Brazil.}
\affiliation{Departamento de F{\'\i}sica Te\'orica,
Universidade do Estado do Rio de Janeiro, Rua S\~ao Francisco Xavier 524, 20550-013  
Rio de Janeiro, Brazil}
\author{Nathan O. Silvano}
\affiliation{Departamento de F{\'\i}sica Te\'orica,
Universidade do Estado do Rio de Janeiro, Rua S\~ao Francisco Xavier 524, 20550-013  
Rio de Janeiro, Brazil}
\author{Daniel G. Barci}
\affiliation{Departamento de F{\'\i}sica Te\'orica,
Universidade do Estado do Rio de Janeiro, Rua S\~ao Francisco Xavier 524, 20550-013  
Rio de Janeiro, Brazil}

\date{\today}

\begin{abstract}
Electronic fluids can display exciting dynamical properties. In particular, due to Landau damping, the collective modes spectrum of an electronic system with multipolar interactions is non-hermitian, and can present non-hermitian degeneracies called {\em exceptional points}. In this work, we want to explore the dynamical properties of these degeneracies using laser control. We show that by using a light pulse, we can control the collective mode spectrum and tune a non-hermitian $\mathcal{PT}$ phase transition in which two exceptional points annihilate each other. At this transition, the gap closes with a cubic root signature, what defines a third order exceptional point. 
\end{abstract}
\maketitle

\section{Introduction}
Order-parameter fluctuations of Fermi liquids are a rich research field from the theoretical and experimental point of view. Although the seminal zero sound excitation, a sound mode propagating in a three-dimensional Fermion liquid, was computed by Landau~\cite{landau1957} and measured in experiments with He$^3$ at low temperatures~\cite{Abel1966,KEEN1963} in the 50's and 60's, this research line regained relevance with the increasing interest in low-dimensional systems~\cite{CastroNeto2020}. Nowadays, there is an interest in electronic systems with multipolar interactions, where the quadrupolar interaction drives nematic phase transitions in a plethora of compounds~\cite{FradKiv2010,Fernandes2014}. The order-parameter fluctuations in these multipolar Fermi liquids represents deformations of the Fermi surface and have been studied in different regimes~\cite{OgKiFr2001, Lawler2006, Aquino-2019, Chubukov2019, Chubukov2020, Sodemann2019, Sankar2022}.

In the Landau Theory of Fermi liquids, each multipolar interaction channel is parametrized by a set of Landau parameters $\{F_\ell\}=\{ F_0, F_1, F_2, \ldots\}$. This class of Fermi liquids possess Pomeranchuk instabilities~\cite{pome58} for strong attractive interactions ($F_\ell < -1$). The Pomeranchuk instabilities induce  spontaneous rotational symmetry breaking producing an stable distortion of the Fermi surface \cite{Lawler2006} such that its shape in the broken phase depends on each angular momentum channel. In Fermionic two-dimensional systems we can write each order-parameter, in the charge sector, as $\mathcal{O}_\ell({\bf q}) = \sum_{\bf k} g_\ell(\theta)\psi^\dagger_{\bf k+q} \psi_{\bf k}$, where $g_\ell(\theta)$ is a form factor that depends on each multipole channel $\ell$, on the polar angle of the Fermi surface $\theta$ and $\psi_{\bf k}$ is a spinless Fermionic field operator.

Interestingly, the dynamics of the order-parameter for $\ell \geqslant 1 $ is non-hermitian. Although the reality of pure quantum non-hermitian Hamiltonians is still under debate, they are widely used in effective descriptions of classical systems with dissipation, where there is control of the energy loss and unique non-hermitian properties can be probed, as for instance the skin effect \cite{Bergholtz-2021}. The field with most developed non-hermitian realizations is photonics \cite{Feng2017, Longhi_2017, Ganainy2018, valagiannopoulos2016pt, Zhiyenbayev2019}, where non-hermiticity allows for more control of wave guides with gain and loss. Moreover, there are other experimental realizations of these systems, as in topoeletrical circuits \cite{Lee2018} and acoustic wave guides \cite{Zhang2021-2}.

There are distinct ways to introduce dissipation in these systems. In Bloch bands, usually one accounts for dissipation by on-site loss or by introducing asymmetrical hopping between sites. In metallic systems, one can introduce dissipation through different quasiparticles scattering rates, a mechanism which have been used in heavy Fermions \cite{Yoshida2018, Yoshida2020}, magnetic systems \cite{Metzner2021, Crippa2023} or disordered systems \cite{Michen2021, Michen2022} for example. In the case of our system of interest, we are not focusing on the quasiparticle bands. Instead, we want to focus on the collective excitations of the electronic liquid, i.e, the order parameter $\mathcal{O}_\ell({\bf q})$ excitation spectrum, where the dissipative nature comes from the energy exchanged between quasiparticles and collective modes, \textit{i.e.}, the Landau dissipation \cite{landau1957, nozieres-1999}, a feature of the electron-electron interaction. 

By considering this dissipation mechanism, we can conclude that effective theories which describe collective modes of metallic systems with multipolar interactions are not closed. Quasiparticles represents the bath while the collective modes represent the system of interest. The consequence is that, not only the collective mode spectrum is complex, but also an exceptional point (EP) appears in the weak attractive region of the spectrum ($F_\ell \lesssim 0$)~\cite{Aquino-2020,Aquino-2021}. This non-hermitian degeneracy induces clear signatures in transport and dynamical quantities, however is still an open question if this non-trivial structure represent some kind of phase transition. This is one of the motivations of the present study.

In this work, we will exploit how to control these collective excitations of an electronic fluid with quadrupolar interactions using ultrafast light. Through a minimal coupling between the Fermions and an external electric field, we construct a local effective theory for the collective mode excitation $\boldsymbol{\phi}$ coupled to a laser beam. The light effect is recognizable  when the electric field work $eE_0L$, where $e$ is the electron charge and $L$ is the material length, is comparable to a single particle-hole energy $v_F q$. We found that the external stimulus generates another EP, which approaches the original one for higher intensity of the beam.  Each EP represents a $\mathcal{PT}$ symmetry breaking point,  or a point  where the collective mode spectrum changes from complex to real to complex again, and both complex gap closes with a square root signature. At a critical value of the laser intensity, the $\mathcal{PT}$ symmetric phase, i.e., the real part of the spectrum shrinks, signaling a higher order $\mathcal{PT}$ symmetric breaking, where the complex gap closes with a cubic root signature, what defines a third order EP \cite{Mandal2021}. 

We characterize the spectrum, which is depicted in Fig. \ref{fig: Complex-spectrum}, by identifying the effective Hamiltonians which describe the band closing points, i.e., the second order EPs and the third order EP. Moreover, we show a dynamical phase diagram in Fig. \ref{fig: phase-diagram} in terms of the quadrupolar coupling constant $F_2$ and the rate $a_0 = eE_0L/v_F q$. We see that the critical lines are nothing else as exceptional lines, i.e., a collection of exceptional points parametrized by the two parameters which represents the real to complex transition. This is a striking consequence of the Landau damping, i.e., it does not introduce complex level crossing in a unique way, but this can generate different structures depending on the number of degrees of freedom involved in the collective mode dynamics.

The paper is organized as follows: In Sec. \ref{sec:model}, we define the Fermionic model in which we will work. In Sec. \ref{sec:effective} we compute the effective action of the quadrupolar order parameter in the random phase approximation. In Sec. \ref{sec:ultrafast} we define the electric field configuration in which we will work and compute the ultrafast limit as an approximation to localize the effective action. In Sec. \ref{sec:collective} we semi analytically compute the collective modes by finding the zeros of the effective Lagrangian and describe the $\mathcal{PT}$ phase transitions. Finally we discuss our results in Sec. \ref{sec:conclusion}.

\section{Model}\label{sec:model}
We are interest in studding Fermions interacting with quadrupolar (nematic) fluctuations.   To do  this, we consider a two-dimensional model of Fermions coupled to a two-component gapless Boson. The Fermionic sector of our model is described by the single band non-interacting Hamiltonian $H_0 = \sum_{\bf r} \psi^\dagger \left(\bf r\right) [\varepsilon\left(\nabla\right) - \mu] \psi(\bf r) $. Moreover, we consider the following coupling
\begin{align}
	H_I = - F_2 \int d^2r\, \psi^\dagger \left(\bf r \right) \boldsymbol{\phi}\left(\bf r\right) \cdot {\bf g}\left(\nabla\right) \psi \left(\bf r\right)
\end{align}
where $F_2$ is a coupling constant, ${\bf g}\left(\bf \nabla\right) = \left(\partial_x^2-\partial_y^2, 2\partial_x\,\partial_y \right)$  is the form factor, and $\boldsymbol{\phi} = \left(\phi_+, \phi_-\right)$ is the two-component field.
	
Each component of the bosonic field couples to the $d_{x^2-y^2}$ and $d_{xy}$ quadrupolar charge densities, which means that $\boldsymbol{\phi}$ is invariant of rotations by $\pi$. Physically, $\boldsymbol{\phi}$ correspond to a collective mode field. Examples of collective modes are plasmons which correspond to the collective excitations of electrons coupled through Coulomb interactions and phonons, the collective fluctuations of lattice systems. Here, we are interested in describing coherent excitations of fermions coupled through quadrupolar excitations. These quadrupolar fluctuations are often called nematic~\cite{FradKiv2010,Fernandes2014}, which take place in materials such as cuprates, pnictides and heavy-fermions. Our model of a two component nematic order parameter is suitable for describing strongly correlated electronic systems with isotropic nematic fluctuations. Recently, there has been experimental evidence of near-isotropic fluctuations of nematic order parameter in iron pnictides~\cite{Ishida2020} or heavy metal compounds~\cite{Ronning2017}. Moreover, another exciting platform is on cold Fermi gases with quadrupolar interaction~\cite{Bhongale2013}.

Due to the fact that we want to explore pure dynamical properties of such bosons, we consider very low temperatures, in which the thermal fluctuations do not contribute to the processes in which we are interested. 

The goal here is to include now an external electric field through minimal coupling, i.e., $\partial_t \to \partial_t +ie\,U$, where $e$ is the Fermionic electric charge and $U \left({\bf r},t\right)$ is the scalar electric potential. To formally handle this external field, we plug the Hamiltonian in a generating functional , in such a way that the action is written as
\begin{align}
	\mathcal{S} = \int dt d^2 r \left[\mathcal{L}_{\text{ferm}} + \mathcal{L}_{\text{nem}}\right], \label{eq: action}
\end{align}
and
\begin{align}
	\mathcal{L}_{\text{ferm}} &=   \psi^\dagger \left(\bf r\right) \left[G_{0}^{-1} + C \left({\bf r}\right) \right] \psi\left(\bf r\right)  \\
	\mathcal{L}_{\text{nem}} &= \frac{i}{4 \,F_2}\,  \vert \boldsymbol{\phi}\vert^2
\end{align}
with
\begin{align}
	C({\bf r}) &= \frac{1}{2}\boldsymbol{\phi} \cdot {\bf g} - e\, U({\bf r},t) \label{eq: C}
\end{align}
and the non-interacting Green function is given by $G_0^{-1} = i\partial_t - \epsilon(\nabla)$. We will treat the electromagnetic field classically, integrate the Fermions and find the saddle-point solution. In this work, we will consider that the electrons are weakly coupled to the electromagnetic field, so the saddle-point solution is not drastically affected. This means that we can perform a perturbative expansion around the original saddle-point solution ${\boldsymbol{\phi}_0}$. Furthermore, we will focus on the isotropic phase, where $\boldsymbol{\phi}_0 \equiv 0$. To not overcomplicate our notation, from now on we will consider $\boldsymbol{\phi}\left({\bf r}\right)$ the fluctuations of the nematic order parameter. In the next section, we will construct the effective action for the fluctuating field $\boldsymbol{\phi}\left({\bf r}\right)$.

\section{Effective Action}\label{sec:effective}
\begin{figure}[h]
	\centering
	\includegraphics[width=0.45\textwidth]{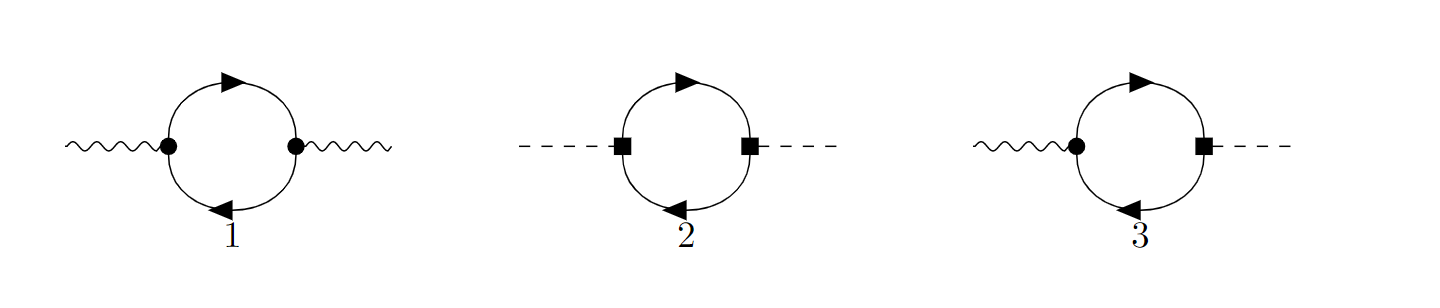}
	\caption{Polarization bubbles in terms of the components of the field $\boldsymbol{\phi}$. The wavy line represents $\phi^+$ and the dashed line $\phi^-$. The filled circle is the vertex for the $\phi^+$ field, $\partial_x^2-\partial_y^2$, while the square vertex is $2\partial_x\partial_y$. Diagrams 1 and 2 give us the diagonal contributions of the dynamical matrix Eq. (\ref{eq: Pi}), while diagram 3 give us the off diagonal contribution.)}
	\label{fig: polarization}
\end{figure}

\begin{figure}[h]
	\centering
	\includegraphics[width=0.45\textwidth]{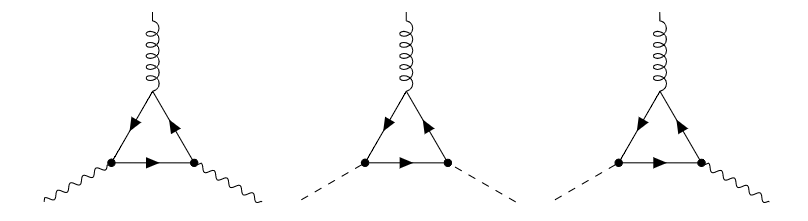}
	\caption{Three leg correction to the Polarization bubble contribution in terms of the components of the field $\boldsymbol{\phi}$. The convention to the vertices and the external legs are the same as in Fig.~(\ref{fig: polarization}).}
	\label{fig: A0-diagrams}
\end{figure}

Performing a Gaussian integration over the Fermionic fields, we find 
\begin{align}
	\label{eq:action1}
	{\cal S} = \int d^2r dt \Bigg\{ \frac{i}{4 F_2}\vert\boldsymbol{\phi}\vert^2 +{\rm Tr\, ln}  \, [1+G_0 C({\bf r})] \Bigg\}.
\end{align}
The main goal of this section is to find and effective action that describes the coupling between the electric potential $U$ and the nematic fluctuations $\boldsymbol{\phi}$ around the saddle-point $\phi_0=0$.  Considering a pertubative calculation in $U$, the first term in the expansion is $U=0$.  Expanding the Logarithm in powers of $C[\boldsymbol{\phi}]$, we find
	\begin{align}
		\label{eq:logexpansion}
		{\rm Tr\, ln}  \, [1+G_0 C({\bf r})] = &	{\rm Tr}  \,[G_0 C({\bf r})] - \frac{1}{2}{\rm Tr}  \,[(G_0 C({\bf r}))^2]\nonumber \\
		&+ \frac{1}{3}{\rm Tr}  \,[(G_0 C({\bf r}))^3]- ...
	\end{align}
Since the saddle-point extremize the action,  all linear terms in the fluctuations cancel.  Thus, the first term in equation (\ref{eq:logexpansion}) is exactly zero.  The first non-vanishing term is the second one and corresponds to the well know RPA polarization bubble~\cite{OgKiFr2001,Lawler2006}.  We depict the polarization bubble in terms of the components of the field $\boldsymbol{\phi}$ in Fig.~(\ref{fig: polarization}).
Computing the integrals over the Fermi surface properly, we arrive at the usual dynamical matrix for the nematic fluctuations over the normal phase, in the small momentum transfer approximation, i.e., ${\bf  q} \ll {\bf k}_F$
\begin{align}
	&{\bf \Pi}_0 \left( {\bf q},\omega\right) = \label{eq: Pi} \\   &=\left(
	\begin{array}{cc}
		\chi_0\left(s\right) + \chi_4\left(s\right)\cos(4\theta)&\chi_4(s) \sin(4\theta)  \\
		\chi_4(s) \sin(4\theta)& \chi_0\left(s\right) - \chi_4\left(s\right)\cos(4\theta)
	\end{array} \right)  \nonumber
\end{align}
where $s=\omega/v_F q$. The functions $\chi_{2\ell}$ are defined as 
\begin{align}
	\chi_{2\ell} (s) = -\delta_{2\ell,0} + \frac{\abs{s}}{\sqrt{s^2 -1}}\left(\abs{s}-\sqrt{s^2 -1}\right)^{2\ell}\Theta(|s|>0),
\end{align}
$\forall \; \ell \in \mathbb{N} $. The behavior of this quantity has been explored in different regimes~\cite{OgKiFr2001, Lawler2006, Chubukov2019, Aquino-2020}. To study its correction due to the coupling with the external field, let us compute the next order of the trace log expansion, which reads, at linear order in $U$
\begin{align}
	\delta S_{A_0} = -e \int_{q_1 q_2}  \boldsymbol{\phi}\left(q_1\right) {\bf \Pi}_{1}\left( q_1, q_2 \right)\boldsymbol{\phi}\left(q_2\right) U \left(-q_1 - q_2\right)
\end{align}
where $q_i = \{ {\bf q}_i, \omega_i \}$ and $\int_{q_1 q_2} = \int \frac{d^2q_1 d^2q_2  d\omega_1 d\omega_2}{\left( 2\pi\right)^6} $. The diagrams that we compute in order to get this contribution are depicted in Fig.~(\ref{fig: A0-diagrams}). 
In the small momentum transfer approximation, i.e., ${\bf q}_1 \ll {\bf k}_F$, ${\bf q}_1 \ll {\bf k}_F$ and ${\bf q}_1 + {\bf q}_2 \ll {\bf k}_F$. For a general configuration of the momentum and frequency of the external field, the correction to the polarization bubble is given by 
\begin{widetext}
	\begin{align}
		{\bf \Pi}_{1}\left( q_1, q_2 \right) = - i N(0) \int_0^{2\pi} \frac{d\theta}{2\pi} \frac{\boldsymbol{g}^2 \left( \theta \right) }  {\omega_1 + \omega_1 - \boldsymbol{v}_F \cdot \left({\bf q}_1 + {\bf q}_2 \right) } \left(\frac{\boldsymbol{v}_F \cdot {\bf q}_1}{\omega_1 - \boldsymbol{v}_F \cdot {\bf q}_1 } - \frac{\boldsymbol{v}_F \cdot {\bf q}_2}{\omega_2 - \boldsymbol{v}_F \cdot {\bf q}_2 } \right), \label{eq: Pi-correction}
	\end{align}
\end{widetext}
where 
\begin{equation}
	\boldsymbol{g}^{2}\left(\theta\right) = \left( \begin{array}{cc}
		\cos^2 \left( 2\theta \right) & 2 \sin \left( 4\theta \right) \\
		2 \sin \left( 4\theta \right) & \sin^2 \left( 2\theta \right)
	\end{array}
	\right) \; , 
\end{equation}
and  $N(0)$ is the density of states at the Fermi surface. 
 
One can solve the angular integration to study the frequency- and momentum-resolved response. For now, let us focus on an external beam with no spatial resolution, ${\bf q}_1 + {\bf q}_2 = 0$. In this limit, Eq.~(\ref{eq: Pi-correction}) can be simplified and, after integration, written in terms of the RPA polarization, Eq.~(\ref{eq: Pi}) 
\begin{align}
	{\bf \Pi}_{1}\left({\bf q}, \omega_1,\omega_2 \right) &= \frac{1}{\omega_1 + \omega_2}  \big({\bf \Pi}_0 ({\bf q}, \omega_2) - {\bf \Pi}_0({\bf q}, \omega_1)\big)
\end{align}
Please note that this momentum $q = |{\bf q}|$ is the momentum of the nematic fluctuation, or, the momentum of a single particle-hole. In terms of the constraint  we  have imposed by choosing a monochromatic electric field, we have ${\bf q}_2 = -{\bf q}_1 = -{\bf q}$. Finally, the total effective action is given by
\begin{align}
	{\cal S}_{\rm eff} &= \frac{N(0)}{2} \int \frac{d^2 q d\omega_1 d\omega_2 }{\left(2\pi\right)^3}\boldsymbol{\phi}\left({\bf q},\omega_1\right) \bigg( \frac{\mathbb{I} }{F_2}\,\delta(\omega_1 +\omega_2) \nonumber \\
	&- {\bf \Pi}({\bf q}, \omega_1,\omega_2)\bigg) \boldsymbol{\phi}(-{\bf q}, \omega_2) \label{eq: Seff}
\end{align}
and
\begin{align}	
	{\bf \Pi}({\bf q}, \omega_1,\omega_2) &=  {\bf \Pi}_0({\bf q}, \omega_1)\delta(\omega_1 +\omega_2) \\ &-{\bf \Pi}_{1}\left({\bf q}, \omega_1,\omega_2\right) U\left({\bf q}=0,-\omega_1-\omega_2\right) \label{eq: Pi-complete} \nonumber
\end{align}
This is our first result.  Eq.~(\ref{eq: Seff}) shows that the external electric field induces retarded dynamics for the order parameter. 

\section{Ultrafast Light Pulse}\label{sec:ultrafast}
Let us choose a proper form for the electric potential  in Eq.~(\ref{eq: Seff}), in order to simplify our analysis. 
In ultrafast experiments, pulsed light fields are used  to control the duration of the external perturbation. With this in mind, let us consider
\begin{equation}
	U\left({\bf r}, t \right) = - E_0 r \cos(\Omega t )\, \text{e}^{-|t|/\tau}\label{eq: A0}
\end{equation}
where $E_0$ is the field amplitude and $\tau$ controls the field duration. We ensure a fast pulse by considering $\tau \ll \Omega^{-1}$, so we disregard the oscillatory part in the analysis. This set a threshold for the central frequency $\Omega$. In this approximation, the width of the envelope is way smaller than the period of oscillation of the field. The advantage of this approximation is that we can work with a huge set of values width of the envelope $\tau$ and the central frequency  $\Omega$. This approximation works for central frequencies from kHZ to THz and widths of ps and fs, as long as we respect the constraint $\tau \,\Omega \ll 1$. Plugging Eq.~(\ref{eq: A0}) into Eq.~(\ref{eq: Seff}), we get
\begin{align}
	&{\cal S}_{\rm eff} =  \int_{{\bf q}, \omega_1,\omega_2} \boldsymbol{\phi}\left({\bf q},\omega_1\right) \left\{\left[{F_2}^{-1} - {\bf \Pi}_0 \left({\bf q},\omega_1\right) \right] \delta\left(\omega_1 + \omega_2\right) \right. \nonumber \\
	&- \left. \delta {\bf \Pi} \left({\bf q},\omega_1,\omega_2\right) \right\} \boldsymbol{\phi}\left(-{\bf q},\omega_2\right)
\end{align}
with $\int_{{\bf q}, \omega_1,\omega_2}  = N(0) \int \frac{d^2 q d\omega_1 d\omega_2}{\left(2\pi\right)^3}$. Moreover, the correction to the polarization bubble is
\begin{align}
	 \delta {\bf \Pi} \left({\bf q},\omega_1,\omega_2\right) = a_0 \frac{v_F q \tau {\bf \Pi}_1\left({\bf q},\omega_1,\omega_2\right)}{1+\tau^2 \left(\omega_1 + \omega_2\right)^2}. \label{eq: delta-Pi}
\end{align}
In Eq.~(\ref{eq: delta-Pi}) we have defined the dimensionless  constant
\begin{align}
	a_0 =  \frac{1}{\sqrt{8\pi}} \frac{e E_0 L}{v_F q}. \label{eq: a0}
\end{align}

Note that for $a_0 \ll 1$, we can disregard the correction to the polarization bubble, and there is no effect of the laser beam on the collective modes. This means that $a_0$ defines $v_F q$ as the natural scale of energy in which we are working. By dimensional analysis we can see that the numerator Eq.~(\ref{eq: a0}) has units of work (i.e., energy), where $eE_0$ is the electric force and $L$ is the typical length of the material sample. So, Eq.~(\ref{eq: a0}) tells us that in order to the electric field be appreciable, the energy of the pulse must be at least of the same order of the particle-hole energy $v_F q$. Of course, for fixed values of $a_0$ we are constraining values of the electric field amplitude $E_0$.

Going back to Eq.~(\ref{eq: delta-Pi}), note that, even though $\tau$ is small, for a sufficient finite value,  $\delta \Pi$ have a narrow peak at $\omega_1 = -\omega_2$.  Imposing  this constraint ,  we can drop the $\omega_2$ dependence of the effective action,
\begin{align}
	{\cal S}_{\rm eff} = \frac{N(0)}{2} \int \frac{d^2 q d\omega}{\left(2\pi\right)^3}  \Phi\left({\bf q},\omega\right) \boldsymbol{\mathcal{L}}_{\rm eff} \left({\bf q},\omega\right) \Phi\left(-{\bf q},-\omega\right) 
\end{align}
and the Lagrangian density reads
\begin{align}
	\boldsymbol{\mathcal{L}}_{\rm eff} \left({\bf q},\omega\right)=  \frac{1}{F_2} \mathbb{I} - {\bf \Pi}_0 \left({\bf q},\omega\right) + a_0 \frac{\partial {\bf \Pi}_0 \left({\bf q},\omega \right)}{\partial \omega}. \label{eq: Leff}
\end{align}
In the next section, we will study the zeros of this Lagrangian density ${\rm det}\left[\boldsymbol{\mathcal{L}}_{\rm eff} \right]= 0$, i.e., the collective mode spectrum.

\section{Collective Mode Spectrum}\label{sec:collective}
\begin{figure*}[htp]
	\centering
	\includegraphics[width=\textwidth]{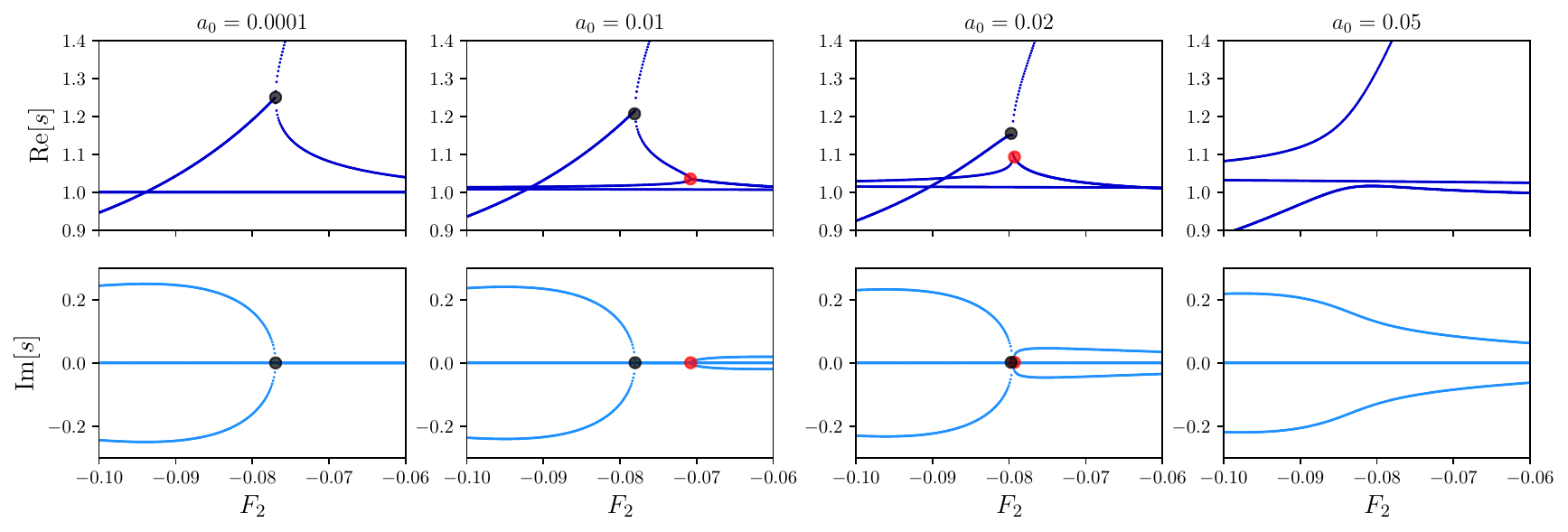}
	\caption{Complex collective mode spectrum for different values of the dimensionless parameter $a_0$. For $a_0 = 0.0001$, we find the spectrum described by Eq.~(\ref{eq: Heff-0}), which presents an exceptional point (black dot) plus two extra collective modes which behaves as free gas excitations. For $a_0=0.01$ a new exceptional point (red dot) that moves from weaker to stronger attraction. For $a_0 = 0.02$ we see both points meeting in parameter space. Finally, for strong enough laser intensity $a_0 = 0.05$, the complex gap is open. }
	\label{fig: Complex-spectrum}
\end{figure*}

Collective excitations are not fully isolated, since they exchange energy with individual quasiparticles. This mechanism of dissipation, called Landau damping, triggers damped modes in the collective mode spectrum. Although this phenomena is not new, the existence of such complex energy levels for the collectives modes is what allows the definition of a complex spectrum. For specific types of electron-electron interactions~\cite{Aquino-2020, Aquino-2021} we see that the Landau damping induce not only damped modes, but {\em exceptional points}. In order to access all this phenomenology, one have to solve the equation
\begin{align}
	{\rm det} \left[\boldsymbol{\mathcal{L}}_{\rm eff}\right] = 0,
\end{align}
which reduces to
\begin{align}
	\prod_{\sigma = \pm} \left[ 1- F_2 \left(\chi_0 + \sigma \chi_2 + a_0 \left(\chi_0^\prime + \sigma \chi_2^\prime \right) \right)\right] = 0. \label{eq: CM-equation}
\end{align}
Recall that $\chi_\ell = \chi_\ell \left(s\right)$, i.e., it is not function of the momentum angle $\theta$. Also, $\chi_\ell^\prime = \partial_s \chi_\ell\left(s\right)$. So, although Eq. (\ref{eq: Leff}) depends on the momentum angle $\theta$, Eq. (\ref{eq: CM-equation}) does not. Each one of the terms that are being multiplied have its own modes of excitations for all $\{\omega,{\bf q}\}$~\cite{Aquino-2019}. Choosing a polarization channel to excite the system, one can simplify Eq.~(\ref{eq: CM-equation}) in order to study only one of the terms~\cite{Aquino-future}. We will choose the term with $\sigma = +$, since its the one that present an exceptional point. So, we will deal with the following equation
\begin{align}
	1- F_2 \left(\chi_0 + \chi_2 + a_0 \left(\chi_0^\prime + \chi_2^\prime \right) \right) = 0 \label{eq: CM-equation-2}
\end{align}
In a previous work~\cite{Aquino-2020}, some of the authors developed the proper treatment for exploring the exceptional point. First, there is no need to solve Eq.~(\ref{eq: CM-equation-2}) for all values of frequency and momenta. In the same spirit that when one study the quasi-homogeneous limit $(\omega \gg v_F q)$, to study quantum critical points, or the quasi-static limit $(\omega \ll v_F q)$, to compute thermodynamical properties, we will study the dynamics of the bosonic order parameter close to the Landau threshold, $\omega \approx v_F q$. The choice of this regime comes from the fact that, while the exceptional point is slightly above the particle-hole continuum, on the other hand, the degeneracy is well separated from the cut $\omega^2 < v_F^2 q^2$. This allows us to perform an expansion of Eq.~(\ref{eq: CM-equation-2}) in powers of $\left(s-1\right)^{1/2}$, arriving at
\begin{align}
	-5 + \frac{2}{\sqrt{s-1}} + 16 \sqrt{s-1} - 2\frac{a_0}{\sqrt{\left(s-1\right)^3}}= \frac{1}{F_2} \label{eq: CM-equation-approx}
\end{align}
Analyzing Eq.~(\ref{eq: CM-equation-approx}), we see that the electric field enters in the dynamics with a highly divergent term, which will induce deep changes in the excitation spectrum. By finding the roots of this equation, we can depict the complex collective mode spectrum in Fig.~(\ref{fig: Complex-spectrum}). 


We see some interesting features on the new spectrum. For $a_0 < 5\times 10^{-3}$, we have the excitation spectrum with an exceptional point plus two degenerate modes, which behaves as non renormalized, i.e., free gas excitations with dispersion given by $\omega = v_F q$.  For $a_0 = 5\times10^{-3}$ the electric field lift one of the degenerate modes and we see a new exceptional point in the spectrum, way closer to the free gas. Using the intensity of the electric field as a tuning parameter, we can control the excitation spectrum and move the singularities across the parameter space. For $ 5\times10^{-3} \leqslant a_0 < 2\times 10^{-2} $ the new exceptional point moves from weak to strong attractive interaction, approaching the original exceptional point. In other words, the exceptional points get closer by increasing $a_0$. At some critical value of $a_0^c = 2\times 10^{-2}$ both exceptional points not only meet, but annihilate each other, opening a gap in the spectrum.  In this way,  the spectrum becomes complex and gapped. 

In other words, perhaps more physically intuitive, the net effect of the electric field is to induced a dissipative mode for very weak attraction,   shrinking the window in $F_2$ space, where dissipation-less propagation is allowed.   This window narrows with increasing $a_0$ up to a point where no more wave propagation is possible without dissipation.   This is the point where a complex gap opens. We identify this behavior as typical of $\mathcal{PT}$ phase transitions. These are characterized by spectra with either real eigenenergies or complex conjugate pairs, in which we call $\mathcal{PT}$ symmetric and broken region, respectively \cite{Bender1998,Bender2002,Bender_2007}. The point in which the real eigenenergies split into complex conjugates, i.e., the point in which the $\mathcal{PT}$ symmetry is broken, is an exceptional point. We can conclude that $a_0$ is the parameter to control the dynamical transition in the collective mode spectrum. The more that we increase $a_0$, the more the $\mathcal{PT}$ symmetric phase shrinks, until it vanishes where the two exceptional points meet. In the following we will write the effective Hamiltonian to describe this spectrum.

\subsection{$\mathcal{PT}$ symmetry breaking}

For $a_0 = 0$, i.e., without any external field, the excitation spectrum has only two levels for $s \approx 1$. This two-level system can be described by the following effective Hamiltonian~\cite{Aquino-2020}:
\begin{align}
	h_{\rm eff} = \left( \begin{array}{cc}
		\epsilon_1 & i w \\
		i w &\epsilon_2
	\end{array}\right)\label{eq: Heff-0}
\end{align}
where $\epsilon_1=(1/25)(27+10F_2)$, $\epsilon_2=(1/25)(25+10F_2)$ and $w=(1/25)\sqrt{20 |F_2|}$ are real positive numbers in the vicinity of the EP. We see that this Hamiltonian is symmetric under $\mathcal{PT}$  transformations, i.e., $\left[\mathcal{PT},h_{\rm eff}\right] = 0$, where $\mathcal{T}$ represents complex conjugation and $\mathcal{P} = \sigma_z$. For weak attractive quadrupolar interaction, this two levels meet at the exceptional point, acquiring an imaginary part for stronger attraction. 

By raising the intensity of the pulse, we completely change the complex spectrum. We induce two extra modes of excitation which for weak intensity, behave just like free gas excitations. As we pointed out in the previous section, we can lift one of these free gas modes and control a $\mathcal{PT}$ symmetry breaking by merging the two exceptional points. Note that they appear for three of the four modes of the spectrum, so this tell us that we can construct a $3\times 3$ effective Hamiltonian to describe the EPs evolution, in fact, near the merging point, we can identify the following Hamiltonian as the correct one describing the spectrum
\begin{align}
	H_{\rm eff} = \left( \begin{array}{ccc}
		\epsilon & -i \left(1+\epsilon \right) & \left(1-i\right)\epsilon \\
		i \left(1 - \epsilon\right) & 0 & \left(1 + i\right) \left( \epsilon - f\right) \\
		-\left(1+i\right)\epsilon & \left(i-1\right) \left(f + \epsilon\right) & -\epsilon
	\end{array}\right)\label{eq: Heff}
\end{align}
with $f = 1.03 + F_2 $ and $\epsilon = \left( a_0 - 1.5 \right)^{-1}$. We can define the parity and time reversal operators that leave this Hamiltonian invariant, $H_{\rm eff} = \mathcal{PT}H_{\rm eff}^*\left(\mathcal{PT}\right)^{-1}$, as the following matrices
\begin{align}
	\mathcal{P} = \left( \begin{array}{ccc}
		1 & 0 & 0 \\
		0 & -1 & 0 \\
		0 & 0 & 1
	\end{array}\right) \quad {\rm and}\quad
	\mathcal{T} = \left( \begin{array}{ccc}
		1 & 0 & 0 \\
		0 & 1 & 0 \\
		0 & 0 & i
	\end{array}\right)
\end{align}
where $\mathcal{T}\mathcal{T}^*=1$.. We can diagonalize $H_{\rm eff}$ using Cardano's methods~\cite{Mandal2021}, so we have three eigenenergies
\begin{align}
	E_1 &= E_+ + E_- \nonumber \\
	E_2&= b E_+ + b^* E_- \\
	E_3&= b^* E_+ + b E_- \nonumber 
	\label{eq: En}
\end{align} 
where $b = \left(-1+i\sqrt{3}\right)/2$ and
\begin{align}
	E_\pm = \sqrt[3]{q \pm \sqrt{q^2+p^3}}
\end{align}
with parameters $q/\epsilon = \left[ 1-2f^2+4\left(f - 1\right)\epsilon +\epsilon^2 \right] /2$ and $p=\left[-1-2f^2+4\epsilon^2\right]/3$. We see that the structure of the eigenenergies is now a cubic root, instead of a square root. This induces a new type of non-hermitian degeneracy in the spectrum, called third order exceptional point (3EP), which is characterized by a complex band closing with $E_{2/3} \sim F_2^{1/3}$, in fact, at this point, $q = p = 0$ . In our complex spectrum, Fig. (\ref{fig: Complex-spectrum}), the 3EP happen when both exceptional point meet and the complex gap completely closes.

Physically, there are differences between a Landau phase transition, such as Pomeranchuk instabilities in this context, or a $\mathcal{PT}$ symmetry breaking. While in the Landau paradigm the system ground state changes during the phase transition, here the $\mathcal{PT}$ symmetry breaking represents a transition in the excited states of the quantum system. This means that during these dynamical instabilities, the system itself is in one single phase. In our case, the disordered normal Fermi liquid phase. 

\begin{figure}[h!]
	\centering
	\includegraphics[width=0.45\textwidth]{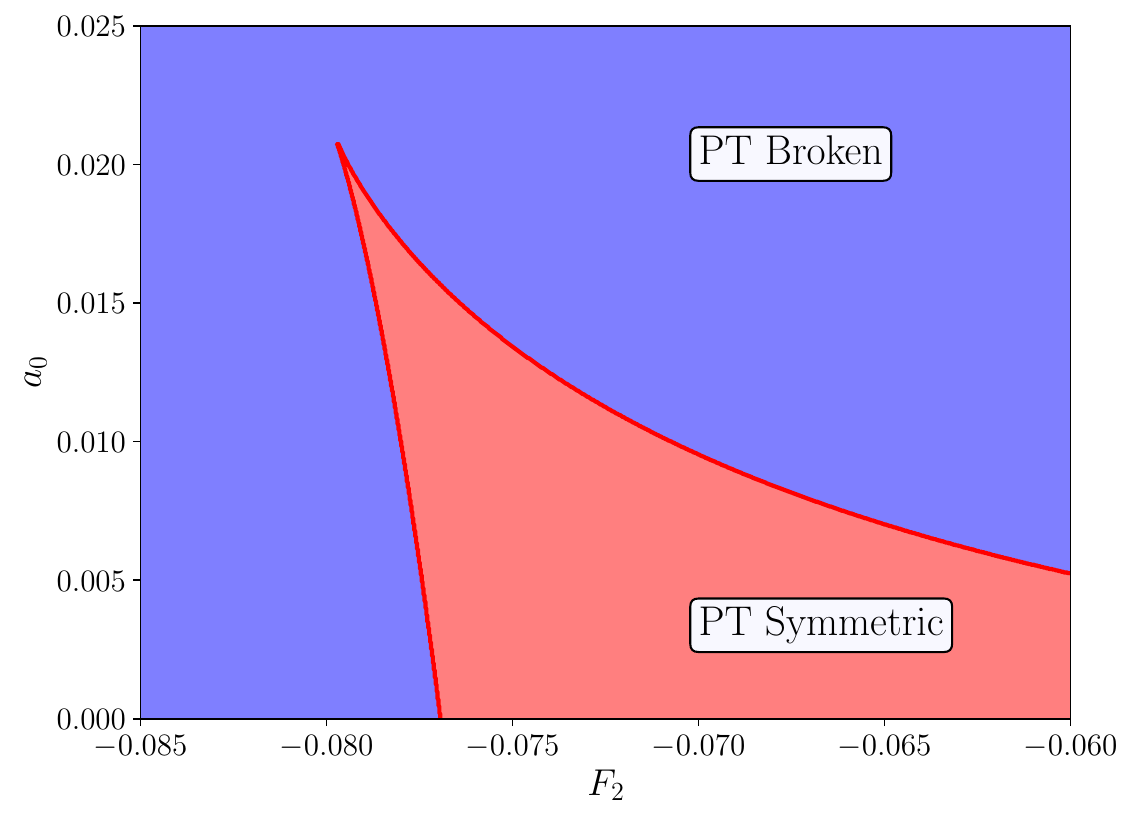}
	\caption{PT Phase diagram. The phase transition lines are characterized as exceptional lines, in which the square root is parametrized by both parameters, $a_0$ and $F_2$. The point in which both lines meet is the third order exceptional point, where the whole gap closes with a cubic root signature. }
	\label{fig: phase-diagram}
\end{figure}

We can construct a dynamical phase diagram with the parameters $\{a_0, F_2\}$. This is shown in Fig. (\ref{fig: phase-diagram}). The red region is given by the values in which $\Im E_\pm =0$, i.e., the spectrum is $\mathcal{PT}$ symmetric and the purple region is given by $\Im E_\pm \neq 0$, i.e., the $\mathcal{PT}$ broken phase. We can see that the two phases are separated by exceptional lines~\cite{Aquino-2020}, as the gap closes as a function of both $a_0$ and $F_2$. 
This characterizes the dynamical phase transitions in the collective mode spectrum induced by the external electric field.


\section{Conclusion}\label{sec:conclusion}

We have presented a study of the collective excitations of a Fermionic fluid with quadrupolar interactions in the normal phase. We analyzed our results in terms of the quadrupolar coupling constant $F_2$ and the dimensionless parameter $a_0$, which is related to the electric field amplitude $E_0$. The first result to note is that the singularity on the excitation spectrum identified previously \cite{Aquino-2020} is not only robust under the action of an external electric field, but the laser makes the complex spectrum richer. It seems that because Landau damping does not uniquely introduce dissipation, the inclusion of more degrees of freedom increases the dimensionality of the non-hermitian spectrum and introduces new structures to it. Previously, it was believed that only the introduction of different coupling constants, as $F_0$ which represents monopole (charge) fluctuations and induce exceptional lines~\cite{Aquino-2019}, or even $F_1$ which represents dipolar interactions~\cite{Aquino-2021,Sodemann2019} and present its exceptional point~\cite{Aquino-2021,Sodemann2019}, that could change the spectrum in a nontrivial way. 

By considering that the electrons are weakly coupled to the external electric field, the saddle point solution for the system without an external field is not drastically changed. Consequently, we compute the fluctuations over the original saddle-point solution and arrive at a non-local action, Eq.~(\ref{eq: Seff}), which describes the collective mode $\boldsymbol{\phi}$ coupled with a pulsed electric field. After this point, we perform two approximations to study the complex spectrum near the exceptional point. First, we consider that the electric field is fast, i.e., the width of the envelope is way smaller than the period of oscillation, which allowed us to localize the action  Eq.~(\ref{eq: Seff}). The other approximation consist in recognizing that the non-hermitian degeneracy appears close to the Landau threshold, so in the same spirit as performing the quasi-homogeneous $(\omega \gg v_F q)$, or the quasi-static limit $(\omega \ll v_F q)$, we perform a series expansion around $\omega \approx v_F q$. Up to these approximations, we get the collective mode equation, Eq.~(\ref{eq: CM-equation-approx}), which we can solve.

There are four solutions of Eq.~(\ref{eq: CM-equation-approx}). Two of them are the same as the collective modes without the electric field, the quadrupolar zero sound and a fast mode that comes from $s\to \infty$ and have a renormalized velocity $v_F^* \gg v_F$~\cite{Aquino-2019}. These two real modes coalesce at strong enough attraction, signaling an $\mathcal{PT}$ symmetry breaking point with a square root type of gap closing. It is worth noting that these modes are almost not changed by the electric field, with the position of the EP being slightly modified. For a nonzero electric field, $a_0 \neq 0$, two new modes are introduced in the spectrum, both behaving as free gas modes. As we increase the electric field intensity, the new EP moves from weak to strong attraction until it meets the former EP. At this point in which both singularities merge, the complex gap closes with a cubic root signature, which defines a third-order exceptional point. 

We characterize these EPs behavior as being a high order $\mathcal{PT}$ phase transition in the excitation spectrum. To do so, we identify that the Hamiltonian Eq.~(\ref{eq: Heff}) describes the third order EP and is $\mathcal{PT}$-symmetric. By finding its spectrum Eq.~(\ref{eq: En}) we can identify the complex level crossing and confirm that the spectrum in Fig.~(\ref{fig: Complex-spectrum}) is indeed described by a non-Hermitian Hamiltonian. Both EPs represents a second order $\mathcal{PT}$ symmetry breaking and the trajectory of the laser-induced EP represents a shrinking of the phase with real eigenvalues. We can see this behavior in the phase diagram Fig.~\ref{fig: phase-diagram}, where we can see both regions as functions of the quadrupolar coupling constant $F_2$ and the dimensionless parameter $a_0$. Both critical lines are nothing else as exceptional lines, i.e., the point in the parameter space where the modes change from real to complex conjugate pairs. 

It would be interesting to explore the extent to which Landau damping can induce complex level crossing, in addition to the well-known damped modes for $s < 1$. One promising platform would be magnetic collective modes, i.e., magnons~\cite{Bonetti2022}. For strongly correlated electron systems, is unavoidable the need to deal with acoustic phonons, which most times are excited dynamically together with zero sound excitations. To this extent, would be interesting to study collective modes of systems with Ising and Potts nematic fluctuations~\cite{Fernandes2014,Liu2021}. On the other hand, optical lattices~\cite{Bhongale2013} seem to be an exciting platform, where there is great control of the interaction strength between the Fermionic gases. In particular, for experimental setups that could probe momentum and frequency multipolar charge excitations, as for instance time resolved resonant inelastic x-ray scattering (RIXS) and momentum-resolved electron energy-loss spectroscopy (M-EELS), we could observe this regime of excitations in strongly correlated materials which present nematic fluctuations. Using data recently obtained from FeSe compounds~\cite{Zeng2010,Wray2008}, where we have a Fermi velocity $v_F$ of order 0.7 eVÅ, we see that by pumping this material with a laser intensity in the KW/cm$^2$ range, momentum resolution of $q = 0.05$ Å$^{-1}$, we could implement our mechanism with central frequencies in the order of kHz and envelopes with width of a few fs. All these values are reasonable for experiments with the current technology at our disposal.

\section{ACKNOWLEDGMENTS}

We would like to acknowledge Rodrigo Pereira, Eduardo Miranda and Rodrigo Arouca for useful discussions. The Brazilian agencies, Fundaç\~ao Carlos Chagas Filho  de Amparo à Pesquisa do Rio de Janeiro (FAPERJ), Fundaç\~ao de Amparo à Pesquisa do Estado de S\~ao Paulo (FAPESP), Conselho Nacional de Desenvolvimento Científico e Tecnológico (CNPq) and Coordenaç\~ao de Aperfeiçoamento de Pessoal de Nível Superior (CAPES) - Finance Code 001, are acknowledged for partial financial support. R.A. is partially supported by a Post-Doctoral Fellowship No. 2023/05765-7 granted by São Paulo Research Foundation (FAPESP), Brazil,  and N.O.S.  is partially supported by a Doctoral Fellowship by CAPES.  


%

\end{document}